\documentclass{article}
\usepackage{amsmath,amssymb}
\usepackage[T1]{fontenc}
\usepackage[utf8]{inputenc}
\usepackage[english]{babel}
\usepackage{color}
\usepackage{latexsym}
\usepackage{enumerate}
\usepackage{tabularx}
\usepackage{url}
\usepackage{hyperref}
\usepackage[affil-it]{authblk}
\usepackage{tikz}
\usepackage{graphicx, caption}
\usepackage{subcaption}

\usepackage{marvosym}

\newcommand{\thickL}{thick $\llcorner$}

\title{Planar Graphs as L-intersection or L-contact graphs\thanks{This
    research is partially supported by the ANR GATO, under contract
    ANR-16-CE40-0009.}}

\author[a]{Daniel Gonçalves}
\author[a]{Lucas Isenmann}
\author[b]{Claire Pennarun}
   
\affil[a]{{\small LIRMM, CNRS \& Univ. de Montpellier, France, \par $\{$daniel.goncalves, lucas.isenmann$\}$@lirmm.fr}}

\affil[b]{{\small LaBRI \& Univ. Bordeaux, UMR 5800, France, claire.pennarun@labri.fr}}

\date{}

\newtheorem{theorem}{Theorem}

\newtheorem{definition}[theorem]{Definition}

\newtheorem{proposition}[theorem]{Proposition}
\newtheorem{lemma}[theorem]{Lemma}

\newenvironment{proof}{\par \noindent \textbf{Proof.} }{\hfill$\Box$\medskip}

\def\epais{0.2}

\newcommand{\lshape}[5]{
\draw (#1,#2) -- (#3,#2);
\draw (#3,#2) -- (#3,#2+\epais);
\draw (#3,#2+\epais) -- (#1+\epais,#2+\epais);
\draw (#1+\epais,#2+\epais) -- (#1+\epais,#4);
\draw (#1+\epais,#4) -- (#1,#4);
\draw (#1,#4) -- (#1,#2);
\draw (#1+\epais,#2+\epais) node[above right] {#5};
}

\begin{document}

\maketitle


\begin{abstract}
  The \emph{$\llcorner$-intersection graphs} are the graphs that have a representation as intersection graphs of axis parallel $\llcorner$ shapes in the plane. A subfamily of
  these graphs are \emph{$\{\llcorner,| ,- \}$-contact graphs} which
  are the contact graphs of axis parallel $\llcorner$, $|$, and $-$ shapes
  in the plane. We prove here two results that were
  conjectured by Chaplick and Ueckerdt in 2013. We show that planar graphs
  are $\llcorner$-intersection graphs, and that triangle-free
  planar graphs are $\{\llcorner,| ,- \}$-contact graphs. 
  These results are 
  obtained by a new and simple decomposition technique for 4-connected triangulations. 
  Our results also provide a much simpler proof of the known fact that planar 
  graphs are segment intersection graphs.

\end{abstract} 

\section{Introduction}

The representation of graphs by contact or intersection of predefined shapes in the plane is a broad subject of research since the work of Koebe on the representation of planar graphs by contacts of circles~\cite{Koebe36}.
In particular, the class of planar graphs has been widely studied in this context.

More formally, assigning a shape $X$ of the plane for each vertex of a graph $G$, we say that $G$ is a $X$-intersection graph if there is a representation of $G$ such that every vertex is assigned to a shape $X$, and two shapes $X_1$, $X_2$ intersect if and only if the vertices they are assigned to are adjacent in $G$.
In the case where the shape $X$ is homeomorphic to segments (resp. discs), a $X$-contact system is a collection of $X$ shapes such that if an intersection occurs between two shapes, then it occurs at one of their endpoints (resp. on their border).
We say that a graph $G$ is a $X$-contact graph if it is the intersection graph of a $X$-contact system.
This definition can be easily generalized if the representation of each vertex is chosen among a family of shapes.

\medskip
The case of shapes that are homeomorphic to a disc has been widely studied; see for example the literature for triangles~\cite{FOR94,GLP12}, homothetic triangles~\cite{KKLS06,Swg17}, axis parallel rectangles~\cite{T84}, squares~\cite{Kenyon04,Schramm93}, hexagons~\cite{GHKK12}, or convex bodies~\cite{S90}.
We here focus on the representation of planar graphs as contact or intersection graphs, where the assigned shapes are segments or polylines in the plane.
The simplest definition of representation of graphs by intersection of curves is the so-called \emph{string}-representation: each vertex is represented by a curve, and two curves intersect if and only if the vertices they represent are adjacent in the graph. It is known that every planar graph has a string-representation~\cite{ehrlich1976intersection}. 
However, this representation may contain pairs of curves that cross any number of times. One may thus take an additional parameter into account, namely the maximal number of crossings of any two of the curves: a \emph{1-string} representation of a graph is a string representation where every two curves intersect at most once. The question of finding a 1-string representation of planar graphs has been solved by Chalopin et al. in the positive~\cite{chalopin2010planar}, and additional parameters are now studied, like order-preserving representations~\cite{BiedlSOFSEM2017}.

\medskip
\emph{Segment intersection graphs} are in turn a specialization of the class of 1-string graphs. It is known that bipartite planar graphs are $\{|,-\}$-contact graphs~\cite{Ben91,Fraysseix94} (i.e. segment contact graphs with vertical or horizontal segments). 
De Castro et al.~\cite{Castro02} showed that triangle-free planar graphs are segment contact graphs with only three different slopes. De Fraysseix and Ossona de Mendez~\cite{Fraysseix07} then proved that a larger class of planar graphs are segment intersection graphs.
Finally, Chalopin and the first author extended this result to general planar graphs~\cite{chalopin2009every}, which was conjectured by Scheinerman in his PhD thesis~\cite{schein84}.

\medskip
A graph is said to be a \emph{VPG-graph} (Vertex-Path-Grid) if it has a contact or intersection representation in which each vertex is a path of vertical and horizontal segments (see \cite{AFjgaa15,cohen2016posets}). Asinowski et al.~\cite{asinowski2012vertex} showed that the class of VPG-graphs is equivalent to the class of graphs admitting a string-representation. They also defined the class \emph{$B_k$-VPG}, which contains all VPG-graphs for which each vertex is represented by a path with at most $k$ bends (see \cite{Fdam16} for the determination of the value of $k$ for some classes of graphs).
It is known that $B_k$-VPG $\not \subseteq$ $B_{k+1}$-VPG, and that the recognition of graphs of $B_k$-VPG is an NP-complete problem~\cite{chaplick2012bend}.
These classes have interesting algorithmic properties (see for example~\cite{Marxiv17} for approximation algorithms for independence and domination problems in $B_1$-VPG graphs), but most of the literature studies their combinatorial properties.

Chaplick et al.~\cite{chaplick2013planar} proved that planar graphs are $B_2$-VPG graphs. This result was recently improved by Biedl and Derka~\cite{BDjocg16}, as they showed that planar graphs have a 1-string $B_2$-VPG representation.

\medskip
Various classes of graphs have been showed to have 1-string $B_1$-VPG representations, such as planar partial 3-trees~\cite{BDarxiv15} and Halin graphs~\cite{francis2016vpg}. Interestingly, it has been showed that the class of segment contact graphs is equivalent to the one of $B_1$-VPG contact graphs~\cite{kobourov2013combi}. This implies in particular that triangle-free planar graphs are $B_1$-VPG contact graphs. This has been improved by Chaplick et al.~\cite{chaplick2013planar} as they showed that triangle-free planar graphs are in fact $\{\llcorner, \ulcorner, |, -\}$-contact graphs (that is without using the shapes $\lrcorner$ and $\urcorner$).

The restriction of $B_1$-VPG to $\llcorner$-intersection or $\llcorner$-contact graphs has been much studied (see for example \cite{Fdam16}) and it has been shown that they are in relation with other structures such as Schnyder realizers, canonical orders or edge labelings~\cite{chaplick2013equilateral}. 
The same authors also proved that the recognition of $\llcorner$-contact graphs can be done in quadratic time, and that this class is equivalent to the one restricted to equilateral $\llcorner$ shapes. Finally, the monotone (or linear) $\llcorner$-contact graphs have been recently studied further, for example in relation with MPT (Max-Point Tolerance) graphs~\cite{2017arXiv170301544R,Cdam17max}.


\paragraph{Our contributions}

The two main results of this paper are the following:

\begin{theorem}
    \label{theorem:contact}
  Every triangle-free planar graph is a $\{\llcorner, | , -\}$-contact graph.
\end{theorem}

\begin{theorem}
    \label{theorem:inter}
  Every planar graph is a $\llcorner$-intersection graph.
\end{theorem}

Both results were conjectured in ~\cite{chaplick2013planar}. Theorem~\ref{theorem:contact} is optimal in the sense that a $\{\llcorner, | , -\}$-contact graph with $n$ vertices has at most $2n-3$ edges, while triangle-free planar graphs may have up to $2n-4$ edges. However, up to our knowledge, the question of whether every triangle-free planar graph is a $\{\llcorner, | \}$-contact graph is open~\footnote{In fact, it has been proven in the Masters thesis (in German) of Björn Kapelle in 2015~\cite[Sec.~3.3]{Kapelle15}, but never published.}.
Theorem~\ref{theorem:inter} implies that planar graphs are in $B_1$-VPG, improving the results of Biedl and Derka~\cite{BDjocg16} stating that planar graphs are in $B_2$-VPG.
Since a $\llcorner$-intersection representation can be turned into a
segment intersection representation~\cite{middendorf1992max}, this also directly provides a rather simple proof of the fact that planar graphs are segment intersection graphs~\cite{chalopin2009every}.
Note that a simple modification of our method can be used to prove that 4-connected planar graphs have a $B_3$-EPG representation~\cite{BP17}, where vertices are represented by paths on a rectangular grid with at most 3 bends, and adjacency is shown by sharing an edge of the grid.

The common ingredient of the two results is what we call \emph{2-sided near-triangulations}.
In Section~\ref{sec:2sided}, we present the 2-sided near-triangulations, allowing us to provide a new decomposition of planar 4-connected triangulations (see~\cite{BDjgaa16} and~\cite{Whitney31} for other decompositions of 4-connected triangulations). This decomposition is simpler than the one provided by Whitney~\cite{Whitney31} that is used in~\cite{chalopin2009every}.
In Section~\ref{sec:CTLCS}, we define thick $\llcorner$-contact systems, (i.e., {$\llcorner$-contact representations in which the shapes have some thickness $\varepsilon$)  with specific properties. We then show that every 2-sided near-triangulation can be represented by such a system. This result is used in Section~\ref{sec:th_contact} to prove Theorem 1. Then in Section~\ref{sec:th_inter} we use 2-sided near-triangulations to prove Theorem 2.

\section{2-sided near-triangulations}\label{sec:2sided}

In this paper we consider plane graphs without loops nor multiple
edges. In a plane graph there is an infinite face, called the \emph{outer face}, and the other faces are called \emph{inner faces}. 
A \emph{near-triangulation} is a 
plane graph such that every inner face is a triangle. 
In a plane graph $G$, a \emph{chord} is an edge not incident to the outer face but that links two vertices of the outer face.
A \emph{separating triangle} of $G$ is a cycle of length three such that both regions delimited by this cycle (the inner and the outer region)
contain some vertices. 
It is well known that a triangulation is
4-connected if and only if it contains no separating triangle.
Given a vertex $v$ on the outer face, the \emph{inner-neighbors} of $v$ are the neighbors of $v$ that are not on the outer face.
We define here $2$-sided near-triangulations (see Figure \ref{fig:example_2sided}) whose structure will be useful in the inductions of the proofs of Theorem~\ref{theorem:inter}, and Theorem~\ref{thm:CTLCS}.

\begin{definition}
  A \emph{2-sided near-triangulation} is a 2-connected near-triangulation $T$ without
  separating triangle and such that going clockwise on its
  outer face, the vertices are denoted $a_1,a_2,\ldots, a_p, b_q,
  \ldots, b_2,b_1$, with $p\ge 1$ and $q\ge 1$, and such that there is
  no chord $a_ia_j$ or $b_ib_j$ (that is an edge $a_ia_j$ or $b_ib_j$
  such that $|i-j| > 1$).
\end{definition}

\begin{figure}[ht]
    \centering
    \includegraphics[scale=0.8]{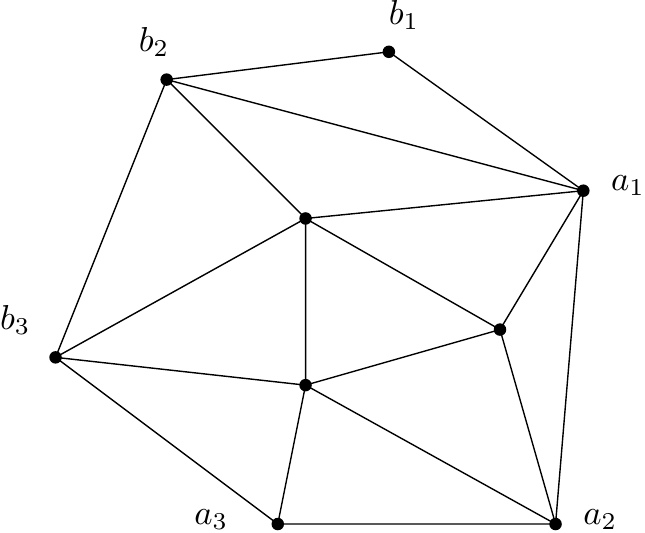}
    \caption{Example of a $2$-sided near-triangulation}
    \label{fig:example_2sided}
\end{figure}

The structure of the 2-sided near-triangulations allows us to describe the following decomposition:

\begin{lemma}~\label{lem:decomp}
  Given a 2-sided near-triangulation $T$ with at least $4$ vertices, one can always perform one of the
  following operations:
  \begin{itemize}
  \item[($a_p$-removal)] This operation applies if $p>1$, if $a_p$ has
    no neighbor $b_i$ with $i<q$, and if none of the inner-neighbors
    of $a_p$ has a neighbor $b_i$ with $i<q$. This operation consists
    in removing $a_p$ from $T$, and in denoting $b_{q+1}, \ldots,
    b_{q+r}$ in anti-clockwise order the new vertices on the outer face, if any. This
    yields a 2-sided near-triangulation $T'$ (see Figure \ref{fig:ap_removal_operation_on_graph}).
    
  \item[($b_q$-removal)] This operation applies if $q>1$, if $b_q$ has
    no neighbor $a_i$ with $i<p$, and if none of the inner-neighbors
    of $b_q$ has a neighbor $a_i$ with $i<p$. This operation consists
    in removing $b_q$ from $T$, and in denoting $a_{p+1}, \ldots,
    a_{p+r}$ in clockwise order the new vertices on the outer face, if any. This
    yields a 2-sided near-triangulation $T'$. 
    This operation is strictly symmetric to the previous one.
    
  \item[(cutting)] This operation applies if $p> 1$ and $q>1$, and if the
    unique common neighbor of $a_p$ and $b_q$, denoted $x$, has a
    neighbor $a_i$ with $i<p$, and a neighbor $b_j$ with $j<q$. If $x$
    has several such neighbors, $i$ and $j$ correspond to the smaller
    possible values. This operation consists in cutting $T$ into three
    2-sided near-triangulations $T'$, $T_a$ and $T_b$
(see Figure \ref{fig:cutting_operation_on_graph}):
    \begin{itemize}
    \item $T'$ is the 2-sided near-triangulation
    contained in the cycle formed by vertices $(a_1,\ldots,a_i,x,b_j,\ldots,b_1)$, and the vertex $x$ is renamed $a_{i+1}$.
    \item $T_a$
    (resp. $T_b$) is the 2-sided near-triangulation contained in the cycle
    $(a_i,\ldots,a_p,x)$ (resp. $(x,b_q,\ldots,b_j)$), where the
    vertex $x$ is denoted $b_1$ (resp. $a_1$).
    \end{itemize}
    
  \end{itemize}
\end{lemma}

\begin{figure}
\centering
    \begin{subfigure}{0.4\textwidth}
        \includegraphics[scale=0.8]{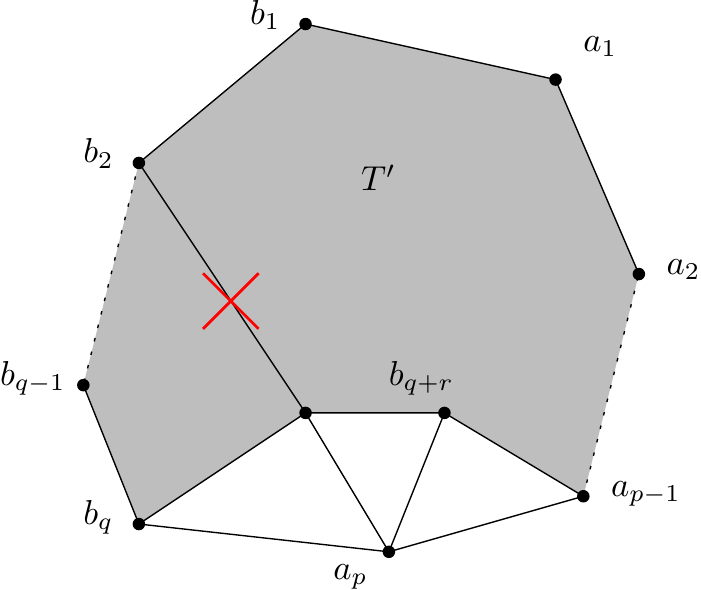}
        \caption{}
        \label{fig:ap_removal_operation_on_graph}
    \end{subfigure}
    \qquad
    \begin{subfigure}{0.4\textwidth}
        \includegraphics[scale=0.8]{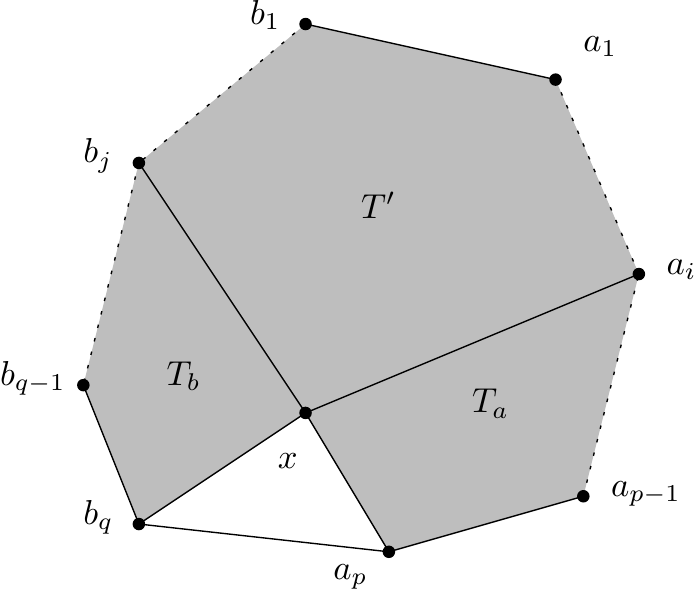}
        \caption{}
        \label{fig:cutting_operation_on_graph}
    \end{subfigure}
    \caption{Illustrations of (a) the $a_p$-removal operation and (b) the cutting operation.}
\end{figure}

\begin{proof}
Suppose that $a_p$ has no neighbor $b_i$ with $i < q$ and none of the
inner-neighbors of $a_p$ has a neighbor $b_i$ with $i < q$.
We denote $b_{q+1}, \ldots , b_{q+r}$ the inner-neighbors of $a_p$ in anti-clockwise order
such that $b_j$ is connected to $b_{j+1}$ for every $q \leq j \leq r$.
Let $T'$ be the graph obtained by removing $a_p$ and its adjacent edges from $T$.
It is clear that $T'$ is a near-triangulation, and that it has no separating triangle (otherwise $T$ would have one too). Furthermore, as there is no chord incident to $a_p$, and as $T'$ has at least three vertices its outer face is bounded by a cycle, and $T'$ is thus 2-connected. 
As $T$ is a $2$-sided near-triangulation, $T'$ has no chord $a_i a_j$, with $i,j< p$, or $b_i b_j$ with $i,j \le q$.
By hypothesis, the inner-neighbors of $a_p$ have no neighbors
$b_k$ with $k < q$, thus there is no chord $b_i b_j$ with $i \le q$ and
$q < j$. There is no chord $b_i b_j$ in $T'$ with $q \le i < j$. 
Otherwise the vertices $a_p$, $b_i$, and $b_j$ would form a triangle with at least one vertex inside, $b_{i+1}$, and at least one vertex outside, $a_{p-1}$: it would be a separating triangle, a contradiction.
Therefore $T'$ is a $2$-sided near-triangulation.

The proof for the $b_q$-removal operation is analogous to the previous
case.

Suppose that we are not in the first case nor in the second one.
Let us first show that $p>1$ and $q>1$. Towards a contradiction, consider that $p=1$. Then as $T$ is 2-connected, it has at least three vertices on the outer face and $q\ge 2$. In such a case one can always perform the $b_q$-removal operation, a contradiction.

Let us now show that $a_p$ is not adjacent to a vertex $b_i$ with $i<q$. Towards a contradiction, consider that $a_p$ is adjacent to a vertex $b_i$ with $i<q$. Then by planarity, $b_q$ (with $q>1$) has no neighbor $a_i$ with $i<p$, and has no inner-neighbor adjacent to a vertex $a_i$ with $i<p$. In such a case one can always perform the $b_q$-removal operation, a contradiction. Symmetrically, we deduce that $b_q$ is not adjacent to a vertex $a_i$ with $i<p$.

Vertices $a_p$ and $b_q$ have one common neighbor $x$ such that $xa_pb_q$ is an inner face. Note that as there is no chord incident to $a_p$ or $b_q$, then $x$ is not on the outer face. They have no other common neighbor $y$, otherwise there would be a separating triangle $ya_pb_q$ (separating $x$ from both vertices $a_1$ and $b_1$). 

As we are not in the first case nor in the second case, we have that $a_p$ (resp. $b_q$) has (at least) one inner-neighbor adjacent to a vertex $b_i$ with $i<q$ (resp. $a_i$ with $i<p$). By planarity, $x$ is the only inner-neighbor of $a_p$ (resp. $b_q$) adjacent to a vertex $b_i$ with $i<q$ (resp. $a_i$ with $i<p$). We can thus apply the cutting operation.

We now show that $T'$, $T_a$ and $T_b$ are 2-sided near-triangulations.
Consider first $T'$.
It is clear that it is a near-triangulation without separating triangles.
It remains to show that there are no chords $a_ia_j$ or $b_ib_j$. By definition of $T'$, the only chord possible would have $x=a_{i+1}$ as an endpoint, but the existence of an edge $xa_k$ with $k<i$ would contradict the minimality of $i$.
Thus $T'$ is a $2$-sided near-triangulation.

By definition, $T_a$ is also a near-triangulation containing no separating triangles.
Moreover, there is no chord $a_k a_l$ with $i \le k \le l-2$ as there are no such chords in $T$.
Therefore $T_a$ is a $2$-sided near-triangulation.
We show in the same way that $T_b$ is a $2$-sided near-triangulation.
\end{proof}

\section{Thick $\{ \llcorner \}$-contact system}
\label{sec:CTLCS}

A \emph{\thickL} is a $\llcorner$ shape where the two
segments are turned into $\varepsilon$-thick rectangles (see
Figure~\ref{fig:thickL}).
Going clockwise around a thick $\llcorner$ from the bottom-right corner, we call its sides
\emph{bottom}, \emph{left}, \emph{top}, \emph{vertical
  interior}, \emph{horizontal interior}, and \emph{right}.

A \thickL~is described by four coordinates $a,b,c,d$ such that $a+\varepsilon < b$ and $c+\varepsilon < d$. It is thus the union of two boxes: $([a,a+\varepsilon] \times [c,d]) \cup ([a,b] \times [c,c+\varepsilon])$. If not specified, the \emph{corner} of a \thickL~denotes its bottom-left corner (with coordinate $(a,c)$). In the rest of the paper, all the \thickL~shapes have the same thickness $\varepsilon$.

\begin{definition}
Given a \thickL-contact system, a thick $\llcorner$ is said \emph{left} if its horizontal interior is free (i.e., does not touch another $\llcorner$) and its left side is not contained in the side of another thick $\llcorner$  (see Figure~\ref{fig:left_bottom_1}). Similarly, a thick $\llcorner$ is said \emph{bottom} if its vertical interior is free and its bottom side is not contained in the side of another thick $\llcorner$ (see Figure~\ref{fig:left_bottom_2}).
\end{definition}

\begin{figure}[ht]
    
    \centering
    \begin{subfigure}[t]{0.3\textwidth}
         \includegraphics[scale=0.8]{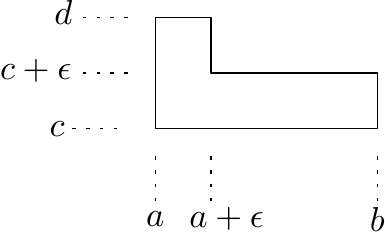}
        \caption{A \thickL}
        \label{fig:thickL}
    \end{subfigure}
    \quad
    \begin{subfigure}[t]{0.3\textwidth}
        \centering
        \includegraphics[scale=0.7]{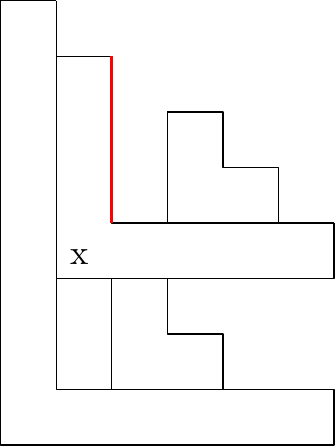}
        \caption{The \thickL~shape representing vertex $x$ is bottom but not left.}
        \label{fig:left_bottom_1}
    \end{subfigure}
    \quad
    \begin{subfigure}[t]{0.3\textwidth}
    \centering
        \includegraphics[scale=0.7]{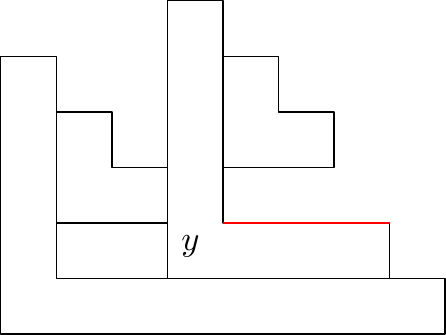}
        \caption{The \thickL~shape representing vertex $y$ is left but not bottom.}
        \label{fig:left_bottom_2}
    \end{subfigure}
    
    \caption{Left and bottom \thickL~shapes.}
    
\end{figure}

\begin{definition}
    A \emph{convenient \thickL-contact system} (CTLCS) is a contact
    system with \thickL~shapes (which implies that the \thickL~shapes
    interiors are disjoint) with a few properties:
    
    \begin{itemize}
    \item Two \thickL~shapes intersect either on exactly one segment or on a
      point (Figure~\ref{fig:CTLCS} lists the allowed ways two
      $\llcorner$ shapes can intersect). If the intersection is a segment, then it must be exactly one side of a \thickL. If the intersection is a point, then it is the bottom right corner of one \thickL~and the top left corner of the other one.
    \item Every thick $\llcorner$ is \emph{bottom} or \emph{left}.
    \end{itemize}
\end{definition}

\begin{figure}[htbp]

\centering

 \begin{tabular}{cccccc}
 \begin{tikzpicture}
  \lshape{1}{1}{2}{2}{}
  \lshape{0}{2}{2}{3}{}
 \end{tikzpicture}
 \hspace{0.2cm}
&
  \begin{tikzpicture}
  \lshape{1}{1}{2}{2}{}
  \lshape{2}{0}{3}{2}{}
 \end{tikzpicture}
 \hspace{0.2cm}
 &
  \begin{tikzpicture}
  \lshape{1}{1}{3}{2}{}
  \lshape{2}{1+\epais}{2.7}{2}{}
 \end{tikzpicture}
 \hspace{0.2cm}
 &
   \begin{tikzpicture}
  \lshape{1}{1}{2}{3}{}
  \lshape{1+\epais}{2}{2}{2.7}{}
 \end{tikzpicture}
 \hspace{0.2cm}
&
 
\begin{tikzpicture}
	\lshape{1}{1}{2}{2}{}
	\lshape{2}{0}{3}{1}{}
\end{tikzpicture}

 \end{tabular}
\caption{Allowed intersections in a CTLCS. From left to right: the intersection is the top, right, bottom, left side of a thick $\llcorner$, and the intersection is a point at the bottom right corner of a thick $\llcorner$ and at the top left corner of a thick $\llcorner$.}
\label{fig:CTLCS}
\end{figure}
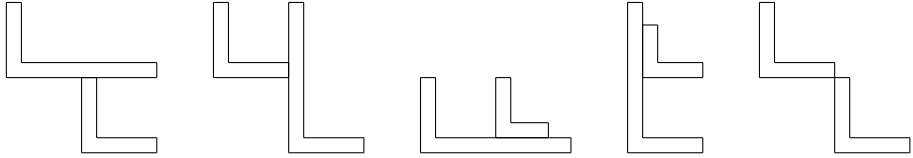

Remark that the removal of any \thickL~still leads to a CTLCS.
This definition implies that in a CTLCS there is no three $\llcorner$ shapes intersecting as in Figure~\ref{fig:CTLCS-bis}.

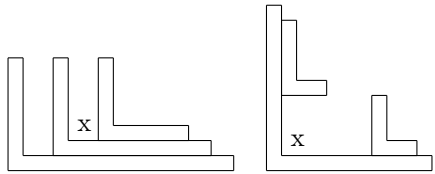
\begin{figure}[htbp]
\begin{center}

\begin{tabular}{cc}
 
\begin{tikzpicture}
 \lshape{0}{0}{3}{1.5}{}
 \lshape{0.6}{0+\epais}{2.7}{1.5}{x}
 \lshape{1.2}{0+2*\epais}{2.4}{1.5}{}
\end{tikzpicture}
&
\begin{tikzpicture}
	\lshape{0}{0}{2.2}{2.2}{x}
	\lshape{0+\epais}{1}{0.8}{2}{}
	\lshape{1.4}{0+\epais}{2}{1}{}
\end{tikzpicture}
\end{tabular}

\end{center}
\caption{Two examples of forbidden configurations in a CTLCS. Here $x$ is not bottom nor left.}
\label{fig:CTLCS-bis}
\end{figure}

We now make a link between CTLCS and 2-sided near-triangulations (See Figure~\ref{fig:CTLCSexample} for an illustration).

\begin{theorem}~\label{thm:CTLCS}
  Every 2-sided near-triangulation can be represented by a CTLCS with the following properties:
      \begin{itemize}
      \item every \thickL~is included in the quadrant $\{ (x,y) : x\ge 0, y \ge 0 \}$,
      \item $a_1$ has the rightmost corner and $b_1$ has the up-most corner,
      \item every vertex $a_i$ is represented by a bottom \thickL~whose corner has
        coordinates $(x,0)$, with $x>0$, and
      \item every vertex $b_i$ is represented by a left \thickL~whose corner has
        coordinates $(0,y)$, with $y>0$.
      \end{itemize}
\end{theorem}

\begin{figure}[!ht]
    \begin{center}
    
    \begin{tabular}{cc}
     
     \includegraphics[scale=0.7]{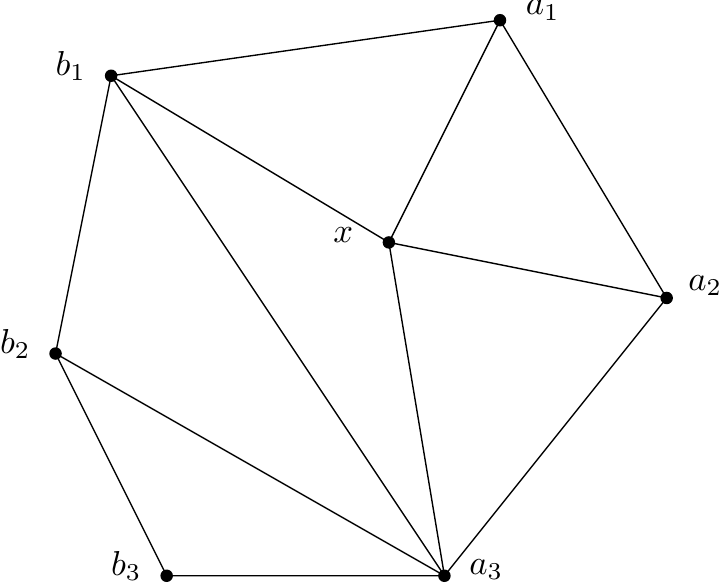}
    
    \hspace{0.5cm}
    &
    \begin{tikzpicture}
    	\lshape{0}{3}{4}{4}{$b_1$}
    	\lshape{0}{2}{1}{3}{$b_2$}	
    	\lshape{0}{1}{1}{2}{$b_3$}	
    	\lshape{1+\epais}{2}{3}{3}{$x$}
    	\lshape{3}{0}{4}{3}{$a_1$}
    	\lshape{2}{0}{3}{2}{$a_2$}
    	\lshape{1}{0}{2}{3}{$a_3$}
    \end{tikzpicture}
    \end{tabular}
    
    \end{center}
    \caption{A 2-sided near-triangulation and (one of) its CTLCS.}
    \label{fig:CTLCSexample}
\end{figure}

\begin{proof}
We proceed by induction on the number of vertices.
The theorem clearly holds for the 2-sided near-triangulation with three
vertices. 
Let $T$ be a 2-sided near-triangulation; it can thus be decomposed using one of the three operations described in Lemma~\ref{lem:decomp}. We go through the three operations successively.

{\bf ($a_p$-removal)} 
Let $T'$ be the 2-sided near-triangulation resulting from an $a_p$-removal operation on $T$. By the induction hypothesis, $T'$ has a CTLCS with the required properties (see Figure~\ref{fig:CTLCS_Tprime}).
We can now modify this CTLCS slightly in order to obtain a CTLCS of $T$ (thus adding a \thickL~corresponding to vertex $a_p$).
Move the corners of the \thickL~corresponding to vertices
$b_{q+1},\ldots, b_{q+r}$ slightly to the right. Since these are left \thickL~shapes, one can do this without modifying the rest of the system.
Then one can add the \thickL~of $a_p$ such that it touches the \thickL~of vertices $b_q$ and $a_{p-1}$ (as depicted
in Figure~\ref{fig:CTLCS_T}). One can easily check that the obtained
system is a CTLCS of $T$ and satisfies all the requirements.

\begin{figure}
    \centering
        
    \begin{subfigure}[t]{0.4\textwidth}
        \includegraphics[scale=0.8]{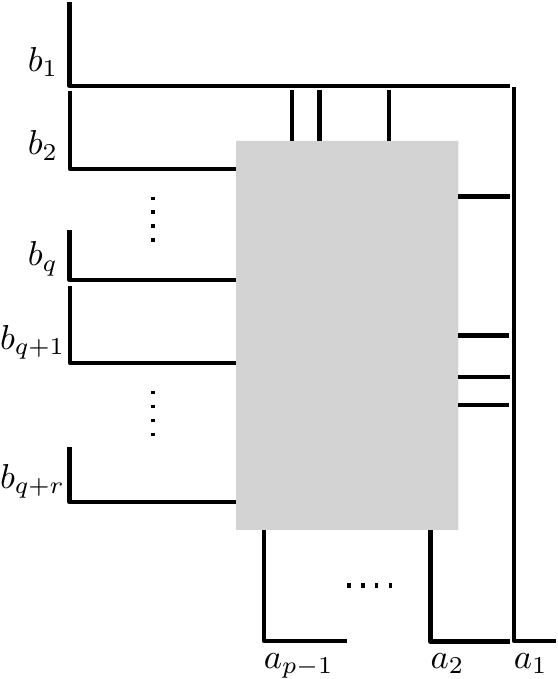} 
        \caption{CTLCS of $T'$}
        \label{fig:CTLCS_Tprime}
    \end{subfigure}
    \quad
    \begin{subfigure}[t]{0.4\textwidth}
        \includegraphics[scale=0.8]{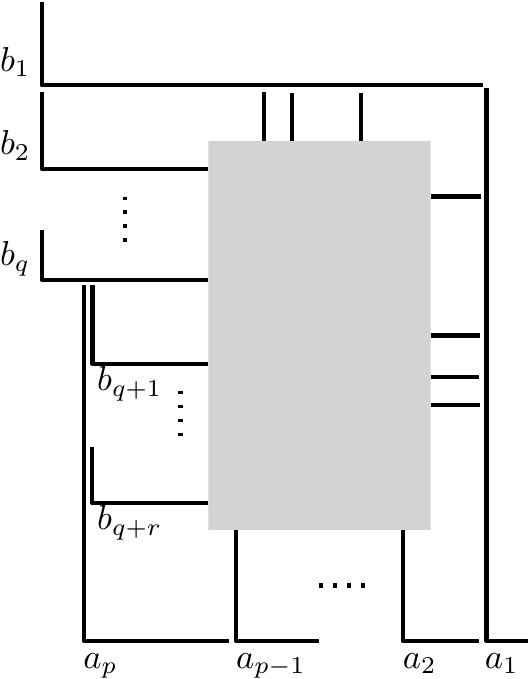}
        \caption{CTLCS of $T$}
        \label{fig:CTLCS_T}
    \end{subfigure}

    \caption{The ($a_p$-removal) operation for a CTLCS. Here, the grey region contains the corners of the inner vertices.}
    \label{fig:thick1}
\end{figure}

{\bf ($b_q$-removal)} This case is strictly symmetric to the previous one.

{\bf (cutting)} Let $T'$, $T_a$ and $T_b$ be the three 2-sided near-triangulations resulting from the cutting operation described in Lemma~\ref{lem:decomp}. By induction hypothesis, each of them has a CTLCS satisfying the requirements of Theorem~\ref{thm:CTLCS}.
Consider the CTLCS of $T'$ (see Figure~\ref{fig:thick2}). Move
the corner of $x=a_{i+1}$ slightly upward. Since $x=a_{i+1}$ is a bottom
vertex, one can do this without modifying the rest of the system. 
Then one can add the CTLCS of $T_a$ below vertex $x$ and the one of $T_b$ on its left (see Figure~\ref{fig:thick2}, bottom). One can easily check that the
obtained system satisfies all the requirements.
\end{proof}

    


\begin{figure}[!ht]
    \centering
    \includegraphics[page=5,scale=0.85]{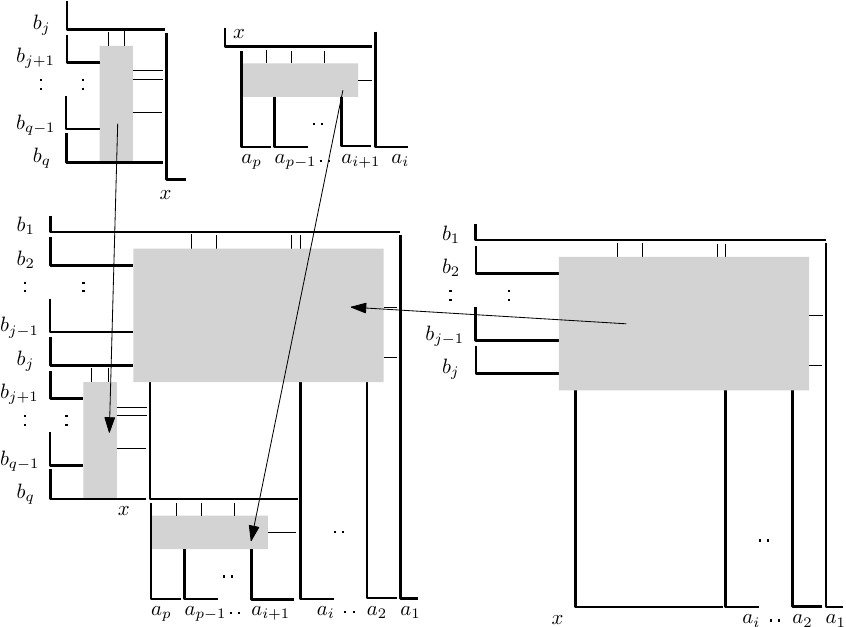}
    \caption{The (cutting) operation for a CTLCS.}
    \label{fig:thick2}
\end{figure}





\section{$\{\llcorner,| ,- \}$-contact systems for triangle-free planar graphs}
\label{sec:th_contact}

We can now use the CTLCS systems to prove Theorem~\ref{theorem:contact}. Recall that a 
\emph{$\{\llcorner,| ,-\}$-contact system} is a contact system with some $\llcorner$, some vertical segments $|$, and some horizontal segments $-$, such that if an intersection occurs between two of these objects, then the intersection is an endpoint of one of the two objects.
We need the following lemma as a tool (it is proved in appendix).

\begin{lemma} \label{lem:triangleFreeTo4Connected}
    For any plane triangle-free graph $G$, there exists a 4-connected triangulation $T$ containing $G$ as an induced subgraph.
\end{lemma}

  
We can now prove Theorem \ref{theorem:contact}, which asserts that every triangle-free planar graph has a $\{\llcorner,| ,-\}$-contact system.

\medskip
\begin{proof}
Consider a triangle-free planar graph $G$.
According to Lemma~\ref{lem:triangleFreeTo4Connected}, there exists a
4-connected triangulation $T$ containing $G$ as an induced subgraph.
As the exterior face of $T$ is a triangle, $T$ is a $2$-sided near-triangulation (denoting $a_1,b_2,b_1$ the three exterior vertices in clockwise order). 
By Theorem~\ref{thm:CTLCS}, $T$ has a
CTLCS and removing every \thickL~corresponding to a 
vertex of $T\setminus G$ leads to a CTLCS of $G$. 

If a \thickL~$x$ has its bottom side included in the horizontal interior side of another \thickL~$y$, then $x$ is bottom, and so does not intersect anyone on its horizontal interior side. Furthermore, $x$ does not intersect anyone on its right side nor on its bottom right corner. Indeed, if there was such an intersection with a \thickL~$z$, then $y$ and $z$ would also intersect, contradicting the fact that $G$ is triangle-free (see Figure~\ref{fig:bottom_triangle_free}). One can thus replace the \thickL~of $x$ by a thick $|$.

\begin{figure}[ht]
    \centering
        \includegraphics[scale=0.9]{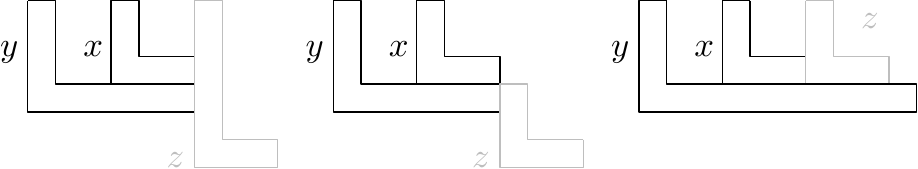}
    \caption{If a \thickL~$x$ has its bottom side included in the horizontal interior of a \thickL~$y$, then $x$ has no intersection with a \thickL~$z$ on its right side and on its bottom right corner.}
    \label{fig:bottom_triangle_free}
\end{figure}

Similarly, if a \thickL~$x$ has its left side included in the vertical interior side of a \thickL~$y$, we can replace the \thickL~of $x$ by a thick $-$.

Note that now the intersections are on small segments, or on a point, between the bottom right corner of a \thickL~or $-$, and the top left corner of a \thickL~or $|$.
Then, we replace each thick $\llcorner$, $|$, and $-$ by thin ones as depicted in Figure~\ref{fig:transfo}. It is clear that we obtain a $\{ \llcorner, |, -\}$-contact system whose contact graph is $G$. This concludes the proof.
An example of the process is shown in Figure~\ref{fig:thick_to_thin}.
\end{proof}

\begin{figure}[ht]
    \centering
    \begin{subfigure}{0.2\textwidth}
        \centering
        \includegraphics[page=1]{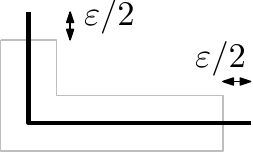}
    \end{subfigure}
    \quad
    \begin{subfigure}{0.2\textwidth}
        \centering
        \includegraphics[page=2]{transfo.pdf}
    \end{subfigure}
    \quad
    \begin{subfigure}{0.2\textwidth}
        \centering
        \includegraphics[page=3]{transfo.pdf}
    \end{subfigure}
    
    \caption{Replacing thick $\llcorner$, $|$, and $-$ by thin ones.}
    \label{fig:transfo}

\end{figure}

\begin{figure}[!ht]
    \centering
    \includegraphics[scale=0.8]{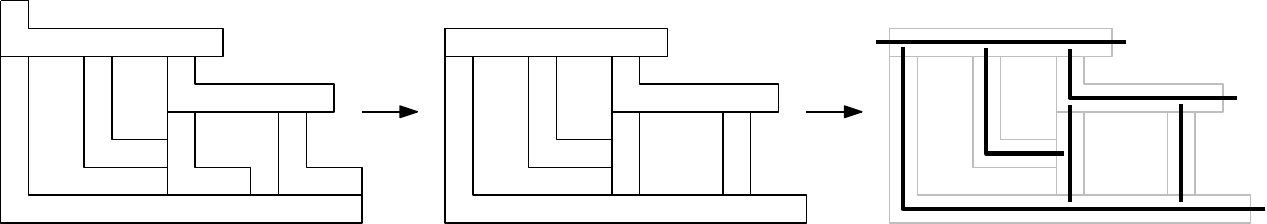}
    \caption{Given a CTLCS of a triangle-free graph $G$, we first replace some thick $\llcorner$ by thick $|$ and thick $-$, and then replace every thick shape by a thin one according to Figure~\ref{fig:transfo}.}
    \label{fig:thick_to_thin}
\end{figure}

\section{The $\llcorner$-intersection systems}
\label{sec:th_inter}

An \emph{$\llcorner$-intersection system} (LIS) is an intersection system of $\llcorner$ shapes where every two $\llcorner$ shapes intersect on at most one point.
Using Theorem~\ref{thm:CTLCS}, one could prove that that every 4-connected triangulation has a LIS. To allow us to work on every triangulation (not only the 4-connected ones) we need to enrich our LISs with the following notion that was introduced in~\cite{Fdam16} under the name of \emph{private region}.

An \emph{anchor} can be seen as a union of three segments, or as the union of two $\llcorner$. It has two \emph{corners}, which correspond to the $\llcorner$ shapes corners. There are two types of anchors.
A \emph{horizontal anchor} is a set $[x_1,x_3] \times y_1 \cup x_1 \times [y_1,y_2] \cup x_2 \times [y_1,y_2]$ where $x_1 < x_2 < x_3$ and $y_1 < y_2$ (see Figure~\ref{fig:anchor_types}a).
The \emph{middle corner} of such a horizontal anchor is defined as the point $(x_2,y_1)$.
A \emph{vertical anchor} is a set $x_1 \times [y_1,y_3] \cup [x_1,x_2] \times y_1 \cup [x_1,x_2] \times y_2$ where $x_1 < x_2 $ and $y_1 < y_2 < y_3$ (see Figure~\ref{fig:anchor_types}b).
The middle corner of such a vertical anchor is defined as the point $(x_1,y_2)$.
Consider a near-triangulation $T$, and any inner face $abc$ of $T$. Given a LIS of $T$, an anchor of $abc$ is an anchor crossing the $\llcorner$ shapes of $a$ $b$ and $c$ and no other $\llcorner$, and such that the middle corner is in the square described by $a$, $b$ and $c$ as depicted in Figure~\ref{fig:anchor_types}.

\begin{figure}[ht]
    \centering
    \begin{subfigure}[t]{0.15\textwidth}
        \centering
        \includegraphics[]{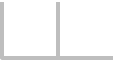}
        \caption{}
    \end{subfigure}
    \qquad
    \begin{subfigure}[t]{0.15\textwidth}
        \centering
        \includegraphics[]{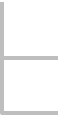}
        \caption{}
    \end{subfigure}
    \qquad
    \begin{subfigure}[t]{0.25\textwidth}
        \centering
        \includegraphics[scale=0.9]{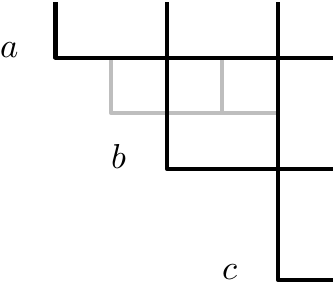}
        \caption{}
    \end{subfigure}
    \qquad
    \begin{subfigure}[t]{0.25\textwidth}
        \includegraphics[scale=0.9]{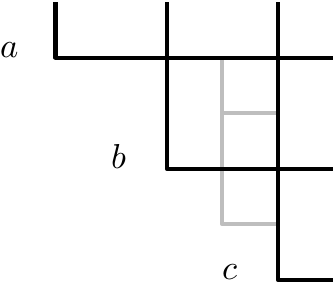}
        \caption{}
    \end{subfigure}
    \caption{The two types of anchors (horizontal and vertical), and the two possible anchors for the $\llcorner$'s of a triangle $abc$.}
    \label{fig:anchor_types}
\end{figure}

\begin{definition}
    A \emph{full $\llcorner$-intersection system} (FLIS) of a near-triangulation $T$ is a LIS of $T$ with an anchor for every (triangular) inner face of $T$, such that the anchors are pairwise non-intersecting.
\end{definition}

\begin{figure}[ht]
    \begin{tabular}{cc}
        \includegraphics[scale=0.6]{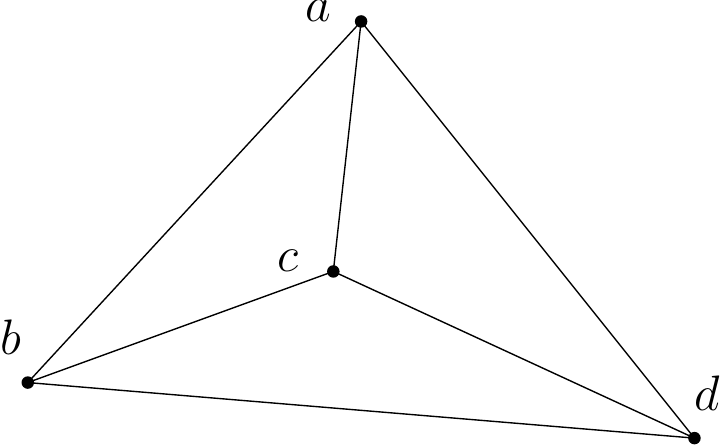}
        &
        \hspace{0.2cm}
        \includegraphics[scale=1]{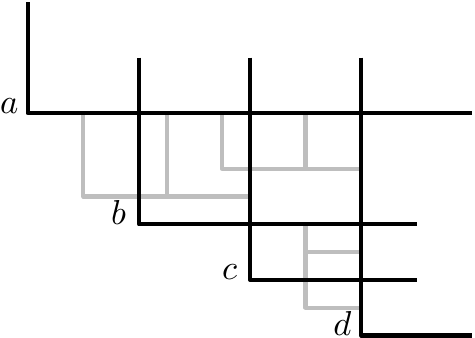}
    \end{tabular}
    \caption{Example of a triangulation and a corresponding FLIS}
    \label{}
\end{figure}

Let us now prove that every 2-sided near-triangulation admits a FLIS.

\begin{proposition}\label{lem:2-sided_FLIS} Every 2-sided near-triangulation has a FLIS such that among the corners of the $\llcorner$ shapes and the anchors:
\begin{itemize}
\item from left to right, the first corners are those of vertices $b_1,b_2,\ldots b_q$ and the last one is the corner of vertex $a_1$, and
\item from bottom to top, the first corners are those of vertices $a_1,a_2,\ldots a_p$ and the last one is the corner of vertex $b_1$.
\end{itemize}
\end{proposition}

As the $\llcorner$ of $a_i$ and $a_{i+1}$ (resp. $b_i$ and $b_{i+1}$) intersect, the FLIS is rather constrained. This is illustrated in Figure~\ref{fig:FLIS-0}, where the grey region contains the corners of the inner vertices, and the corners of the anchors.
  
\begin{figure}[!ht]
    \centering
    \includegraphics[scale=0.7]{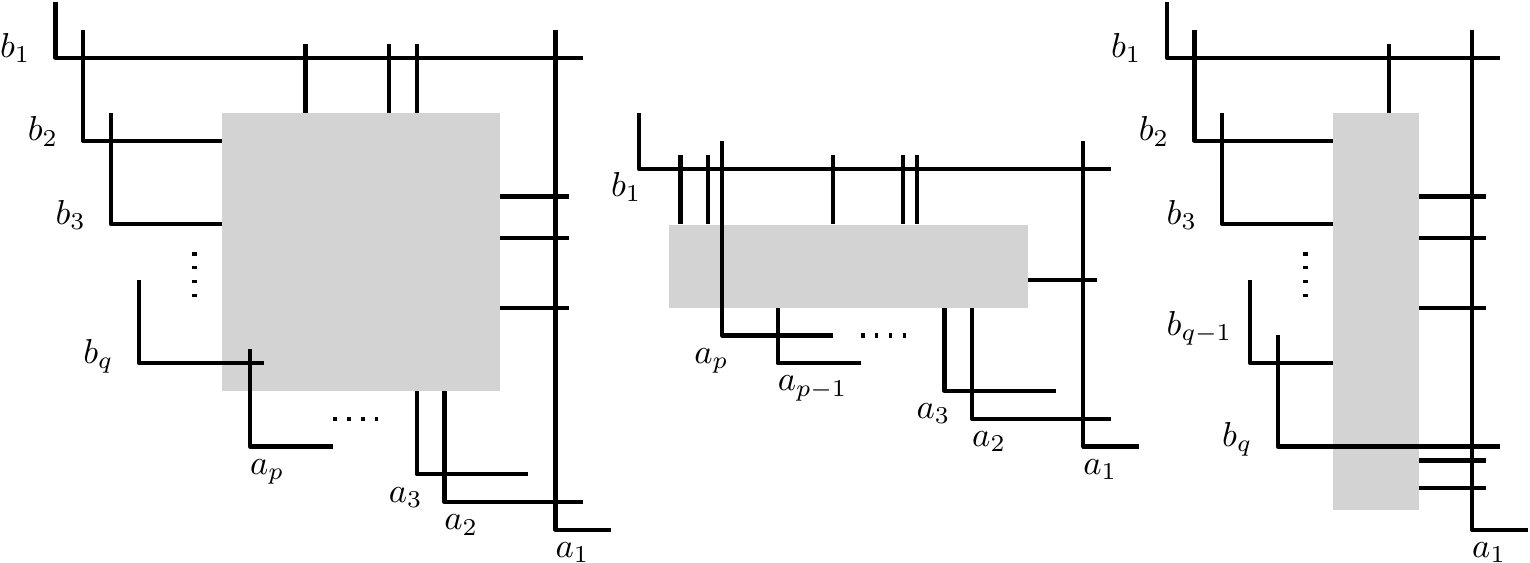}
    \caption{Illustration of Proposition~\ref{lem:2-sided_FLIS} when $p>1$ and $q>1$, when $p=1$ and $q>1$, and when $p>1$ and $q=1$.}
    \label{fig:FLIS-0}
\end{figure}

\medskip
\begin{proof}
We proceed by induction on the number of vertices.

The result clearly holds for the 2-sided near-triangulation with three
vertices, whatever $p=1$ and $q=2$, or $p=2$ and $q=1$. Let $T$ be a 2-sided near-triangulation with at least four vertices.
By Lemma~\ref{lem:decomp} we consider one of the following operations on $T$:

{\bf ($a_p$-removal)} Consider the FLIS of $T'$ obtained by induction
and see in Figure~\ref{fig:FLIS-1} how one can add a $\llcorner$ for $a_p$ and an anchor for each inner face $a_pb_jb_{j+1}$ with $q\le j <q+r$ and for the inner face $a_p a_{p-1} b_{q+r}$. One can easily check that the obtained system verifies all the requirements of Proposition~\ref{lem:2-sided_FLIS}.

\begin{figure}[!ht]
    \centering
        \includegraphics[scale=0.7]{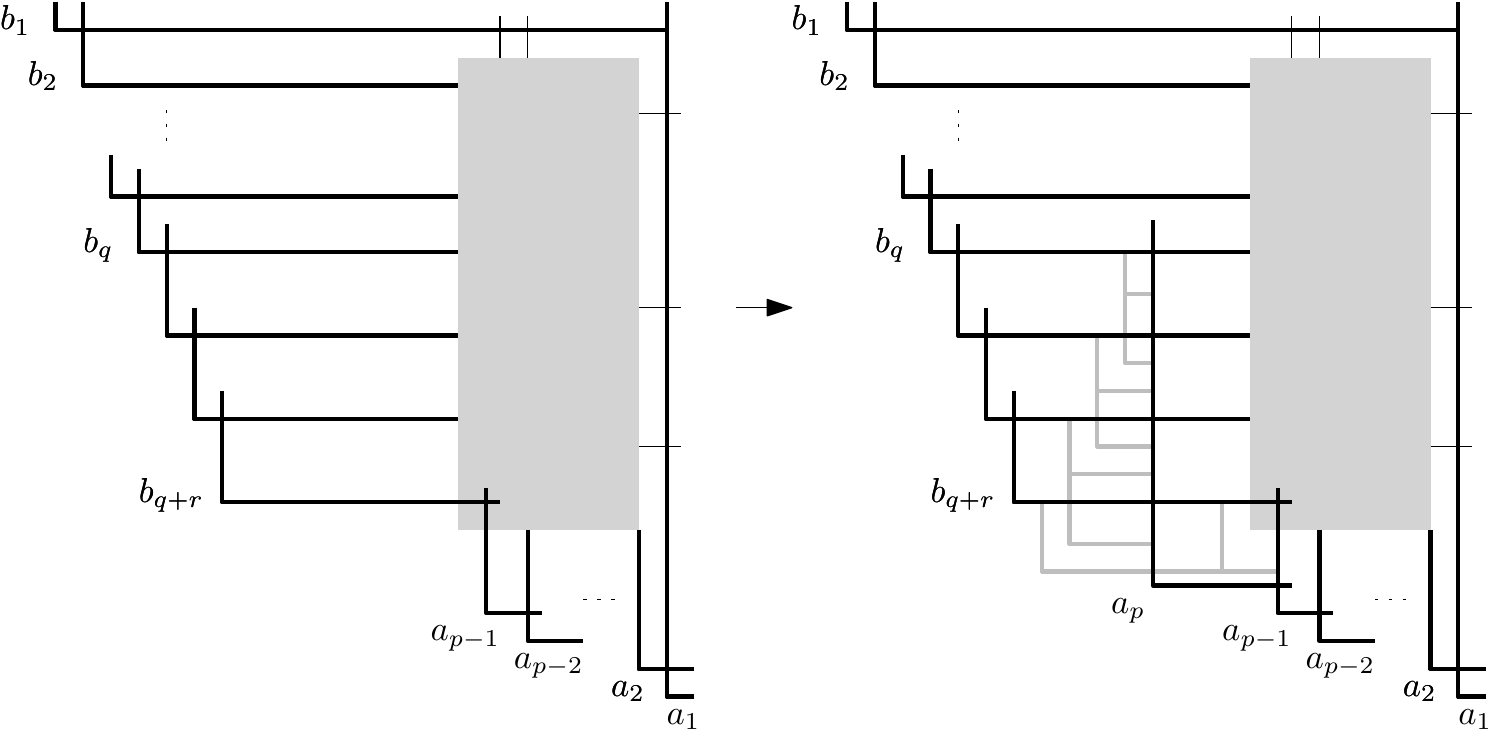}
    \caption{The ($a_p$-removal) operation.}
    \label{fig:FLIS-1}
\end{figure}

\begin{figure}[!ht]
    \centering
    \includegraphics[scale=0.8]{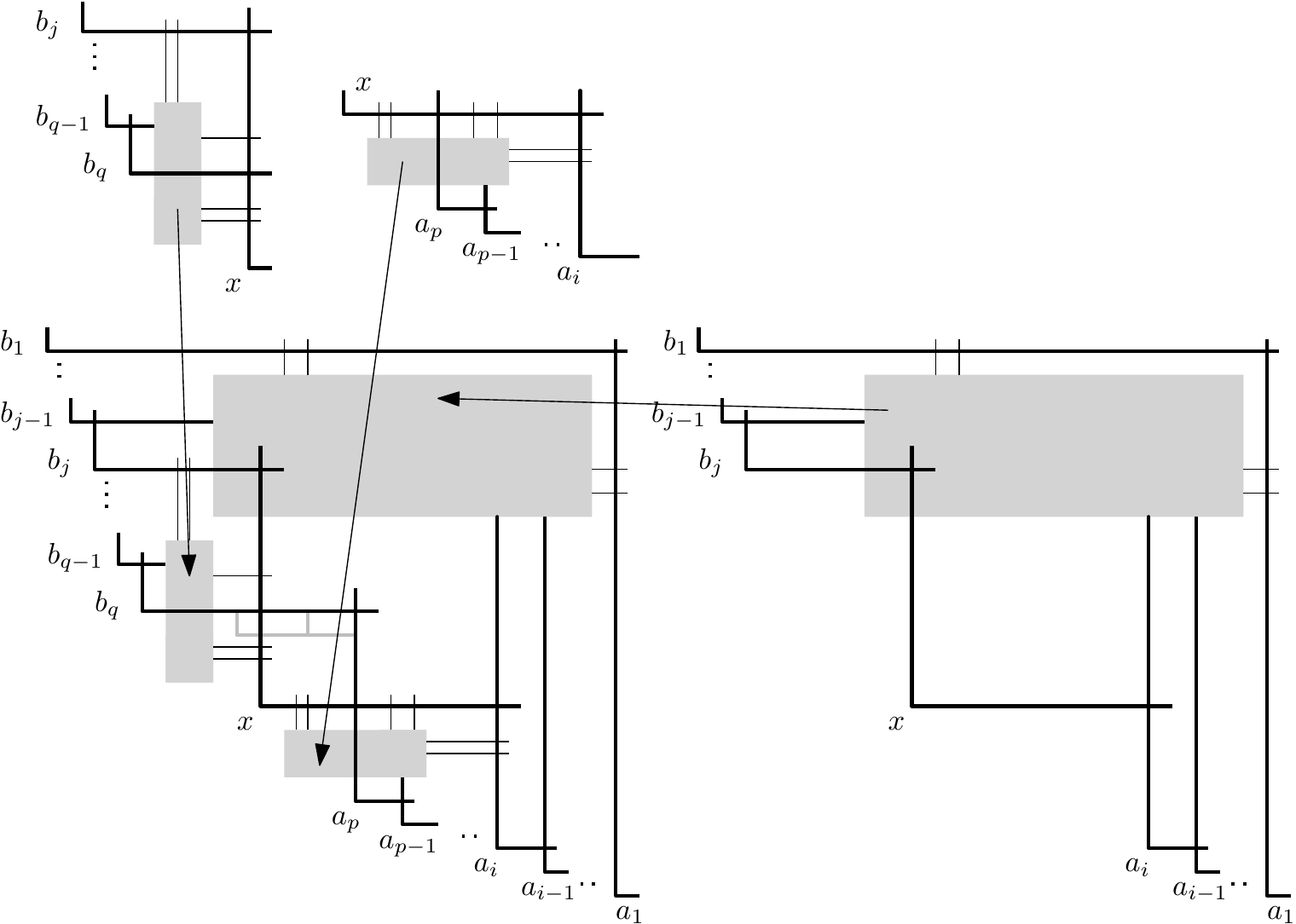}
    \caption{The (cutting) operation.}
    \label{fig:FLIS-2}
\end{figure}

{\bf ($b_q$-removal)} This case is symmetric to the previous one.

{\bf (cutting)} Consider the FLISs of $T'$, $T_a$ and $T_b$. 
Figure~\ref{fig:FLIS-2} depicts how to combine them, and how to add an anchor for $xa_pb_q$, in order to get the FLIS of $T$. One can easily check that the obtained system verifies all the requirements of Proposition~\ref{lem:2-sided_FLIS}.
\end{proof}

We now prove Theorem~\ref{theorem:inter} which asserts that every planar graph is a  $\llcorner$-intersection graph. 
It is well known that every planar graph is an induced subgraph of some triangulation (see~\cite{chalopin2010planar} for a proof similar to the one of Lemma~\ref{lem:triangleFreeTo4Connected}).
Thus, given a planar graph $G$, one can build a triangulation $T$ whose $G$ is an induced subgraph. If one can create a FLIS of $T$, then it remains to remove the $\llcorner$ shapes assigned to vertices of $T \setminus G$ along with the anchors in order to get a $\llcorner$-representation of $G$.
In order to prove Theorem~\ref{theorem:inter}, we thus only need to show that every triangulation admits a FLIS. 

\begin{proposition}~\label{lemma:planar-L-inter}
  Every triangulation $T$ with outer-vertices $x,y, z$ has a FLIS such that among the corners of the $\llcorner$ shapes and the anchors:
  \begin{itemize}
  \item the corner of $x$ is the upmost and leftmost,
  \item the corner of $y$ is the second leftmost, and
  \item the corner of $z$ is the bottom-most and rightmost.
  \end{itemize}
\end{proposition}

\begin{figure}
    \centering
    \begin{subfigure}{0.4\textwidth}
        \includegraphics[scale=0.8]{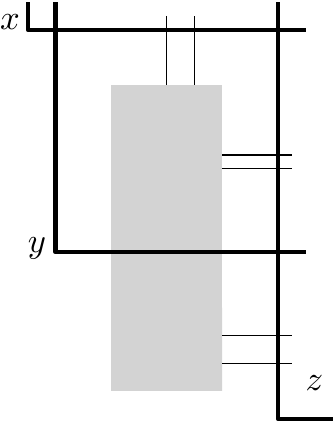}
    \end{subfigure}
    \quad
    \begin{subfigure}{0.4\textwidth}
        \includegraphics[scale=0.8]{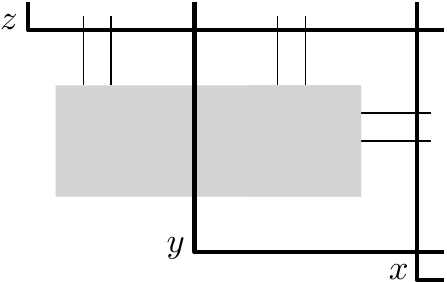}
    \end{subfigure}
    
    \caption{Illustration of Proposition~\ref{lemma:planar-L-inter}, and the FLIS obtained after reflection with respect to a line of slope $1$.
    }    
    \label{fig:thm-FLIS}
\end{figure}


Note that in this proposition there is no constraint on $x, y, z$, so by renaming the outer vertices, other FLISs can be obtained. Another way to obtain more FLISs is by applying a reflection with respect to a line of slope $1$. In such FLIS (see Figure~\ref{fig:thm-FLIS}) among the corners of the $\llcorner$ shapes and the anchors:
  \begin{itemize}
  \item the corner of $x$ is the bottom-most and rightmost,
  \item the corner of $y$ is the second bottom-most, and
  \item the corner of $z$ is the upmost and leftmost.
  \end{itemize}

\begin{proof}
We proceed by induction on the number of vertices in $T$.
Let $T$ be a triangulation with outer vertices $x,y,z$.

If $T$ is 4-connected, then it is also a 2-sided near-triangulation. By Proposition~\ref{lem:2-sided_FLIS} and by renaming the outer-vertices $x$ to $b_1$, $y$ to $b_2$ and $z$ to $a_1$, $T$ has a FLIS with the required properties. 

If $T$ is not 4-connected, then it has a separating triangle formed by vertices $a$, $b$ and $c$.
We note $T_{in}$ and $T_{out}$ the triangulations obtained from $T$ by removing the vertices outside and inside $abc$ respectively.

By the induction hypothesis, $T_{out}$ has a FLIS verifying Proposition~\ref{lemma:planar-L-inter} (considering the outer vertices to be $x,y,z$ in the same order).
Without loss of generality we can suppose that the $\llcorner$ shapes of $a$, $b$ and $c$ appear in the following order: the upmost and leftmost is $b$, the second leftmost is $c$ and the bottom-most is $a$. There are two cases according to the type of the anchor of the inner face $abc$.

If the anchor of $abc$ in the FLIS of $T_{out}$ is vertical (see Figure~\ref{fig:in_out_case1_Tout}), then applying the induction hypothesis on $T_{in}$ with $b,c,a$ as outer vertices considered in that order, $T_{in}$ has a FLIS as depicted on the Figure~\ref{fig:in_out_case1_Tin}.
Figure~\ref{fig:in_out_case1_inclusion} depicts how to include the FLIS of $T_{in} \setminus\{a,b,c\}$ in the close neighborhood of the anchor of $abc$.
As $abc$ is not a face of $T$, the close neighborhood of its anchor is indeed available for this operation.

\begin{figure}[ht]
\centering
    \begin{subfigure}[t]{0.3\textwidth}
        \centering
        \includegraphics[page=2]{sep_triang_case1.pdf}
        \caption{The vertical anchor of $abc$ in the FLIS of $T_{out}$}
        \label{fig:in_out_case1_Tout}
    \end{subfigure}
    \qquad
    \begin{subfigure}[t]{0.26\textwidth}
        \centering
        \includegraphics[page=1]{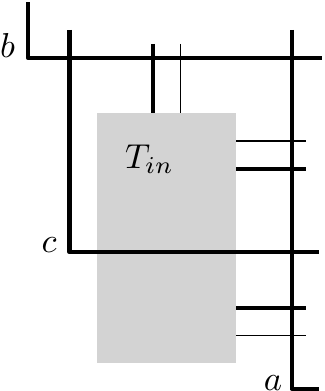}
        \caption{The FLIS of $T_{in}$}
        \label{fig:in_out_case1_Tin}
    \end{subfigure}
    \qquad
    \begin{subfigure}[t]{0.26\textwidth}
        \centering
        \includegraphics[page=3]{sep_triang_case1.pdf}
        \caption{The inclusion of the FLIS of $T_{in}$ in the FLIS of $T_{out}$}
        \label{fig:in_out_case1_inclusion}
    \end{subfigure}

    \caption{FLIS inclusion in the case of a vertical anchor}

\end{figure}

Now suppose that the anchor of $abc$ in the FLIS of $T_{out}$ is horizontal (see Figure \ref{fig:in_out_case2_Tout}).
By application of the induction hypothesis on $T_{in}$ with $a,c,b$ as outer vertices considered in that order, then $T_{in}$ has a FLIS as depicted on the Figure~\ref{fig:in_out_case2_Tin}.
By a reflection of slope $1$, $T_{in}$ has a FLIS such that $b$ is the up-most and left-most, $c$ is the second left-most and $a$ is bottom-most (see Figure~\ref{fig:in_out_case2_Tin_reflected}).
Similarly to the previous case, we include this last FLIS of $T_{in} \setminus \{a,b,c\}$ in the one from $T_{out}$ (see Figure~\ref{fig:in_out_case2_inclusion}). As $T_{in}$ and $T_{out}$ cover $T$, and intersect only on the triangle $abc$, and as every inner face of $T$ is an inner face in $T_{in}$ or in $T_{out}$, these constructions clearly verify Proposition~\ref{lemma:planar-L-inter}.
This concludes the proof of the proposition.
\end{proof}

\begin{figure}[htbp!]
\centering
    \begin{subfigure}[t]{0.4\textwidth}
        \centering
        \includegraphics[page=1]{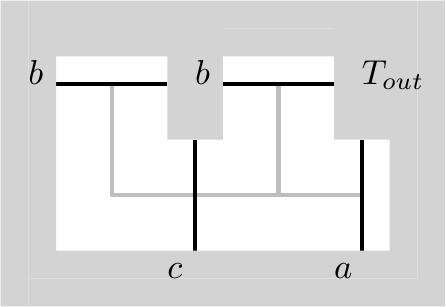}
        \caption{The horizontal anchor of $abc$ in the FLIS of $T_{out}$}
        \label{fig:in_out_case2_Tout}
    \end{subfigure}
    \qquad
    \begin{subfigure}[t]{0.4\textwidth}
        \centering
        \includegraphics[page=2]{sep_triang_case2.pdf}
        \caption{The FLIS of $T_{in}$}
        \label{fig:in_out_case2_Tin}
    \end{subfigure}
        \qquad
    \begin{subfigure}[t]{0.4\textwidth}
        \centering
        \includegraphics[page=3]{sep_triang_case2.pdf}
        \caption{The reflected FLIS of $T_{in}$}
        \label{fig:in_out_case2_Tin_reflected}
    \end{subfigure}
    \qquad
    \begin{subfigure}[t]{0.4\textwidth}
        \centering
        \includegraphics[page=4]{sep_triang_case2.pdf}
        \caption{The inclusion of the FLIS of $T_{in}$ in the FLIS of $T_{out}$}
        \label{fig:in_out_case2_inclusion}
    \end{subfigure}
    
    \caption{FLIS inclusion in the case of a horizontal anchor}
    
\end{figure}






\appendix{}

\section{From triangle-free planar graphs to 4-connected triangulations}

We here prove Lemma~\ref{lem:triangleFreeTo4Connected}.

\begin{figure}[ht]
\centering
\includegraphics[scale=0.9]{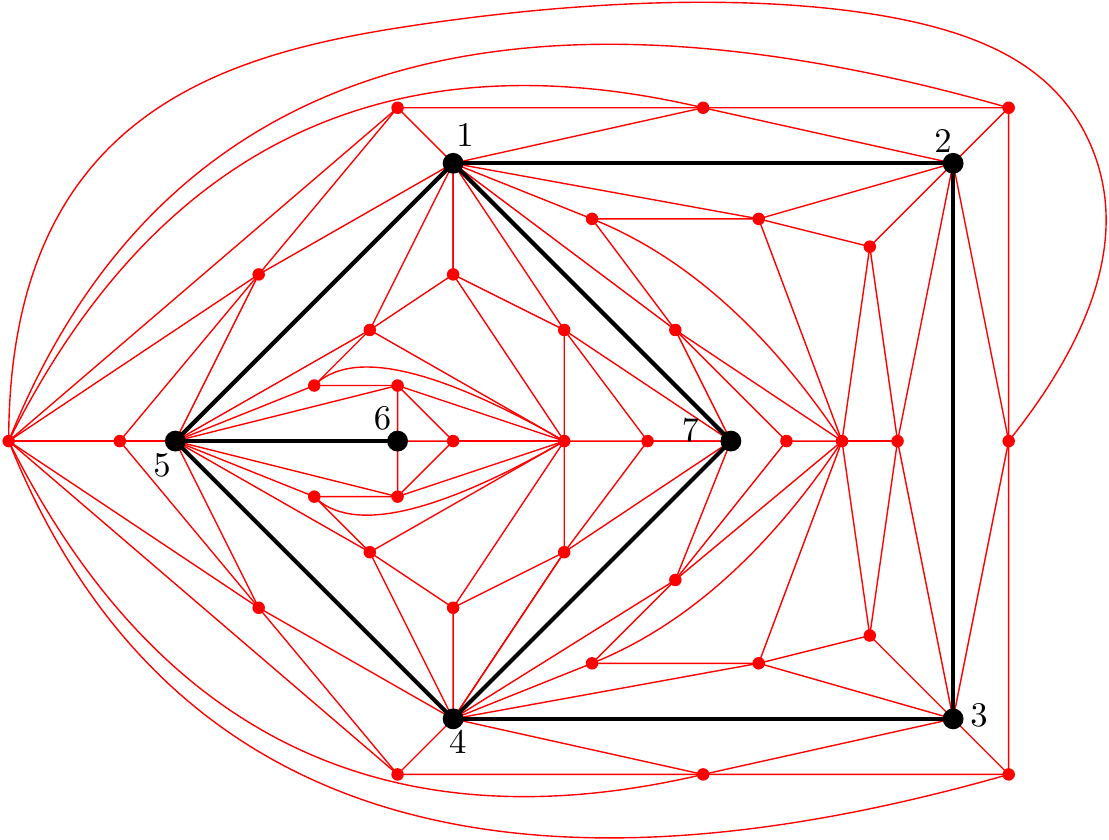}
\caption{A planar triangle-free graph $G$ (in black) and a $4$-connected near-triangulation containing it as an induced subgraph (adding red vertices and edges). The boundary lists of the two inner faces of $G$
are respectively 
$\{1,(1,2),2,(2,3),3,(3,4),4,(4,7),7,(7,1)\}$,
$\{1,(1,7),7,(7,4),4,(4,5),5,(5,6),6,(6,5),5,(5,1)\}$
The outer face is
$\{1,(1,2),2,(2,3),3,(3,4),4,(4,5),5,(5,1)\}$.}
\label{fig:tfreeTo4conn}
\end{figure}

\begin{proof}
The main idea of the construction of $T$ is to insert vertices and edges in every face of $G$ (even for the exterior face).

For the sake of clarity, vertices of $G$ are said \emph{black} and vertices of $T \setminus G$ are said \emph{red}. 
The new graph $T$ contains $G$ as an induced subgraph, along with other vertices and edges. More precisely, for every face of $G$, let $P = \{v_0,e_0,v_1,e_1,\ldots\}$ be the list of vertices and edges along the face boundary (see Figure~\ref{fig:tfreeTo4conn}), where $e_i$ is the edge between vertices $v_i$ and $v_{i+1}$; there can be repetitions of vertices or edges. 
For each face of $G$, given the list $P$, the graph $T$ contains a vertex $v'_i$ for each vertex $v_i$, a vertex $e'_i$ for each edge $e_i$, and an additionnal vertex $t$.
Each vertex $v'_i$ is connected to $e'_i$ and $e'_{i+1}$ (with subscripts addition done modulo the size of the face), each vertex $v_i$ is connected to $v'_i$, $e'_{i-1}$ and $e'_{i}$,
and the vertex $t$ is connected to all vertices $v'_i$ and $e'_i$ (see Figures~\ref{fig:tfreeTo4conn} and \ref{fig:tf24c_gadget} for examples).


\medskip
The new graph $T$ is a triangulation, and we now show that it is 4-connected, i.e., has no separating triangle.
    Suppose that there is a separating triangle in the new graph. There are four cases depending on the colors of the edges of this triangle:
    
    \begin{itemize}
    \item The separating triangle contains three black edges.
    It is impossible since $G$ is triangle-free.
    
    \item The separating triangle contains exactly one red edge.
    One of its endpoints must be a red vertex.
    But a red vertex is adjacent to only red edges, a contradiction.
    
    \item The separating triangle contains exactly two red edges. Then their common endpoint is a red vertex, and the triangle is made of two vertices $v_i$ and $v_{i+1}$, together with the vertex $e'_i$. All these triangles are faces, a contradiction. 
    
    \item The separating triangle contains three red edges.
    Since for each face, the red vertices (vertices $v'_i$, $e'_i$ and $t$) induce a wheel graph centered on $t$, with at least $8$ peripheral vertices (vertices $v'_i$ and $e'_i$), this separating triangle has at least one black vertex. As two adjacent black vertices are linked by a black edge, this separating triangle has exactly one black vertex. As the two red vertices are two adjacent $v'_i$ or $e'_j$ vertices, we have that those are $v'_i$ and $e'_j$, for some $i$ and for $j=i$ or for $j=i+1$. Such a triangle is not separating, a contradiction.
    \end{itemize}
    This concludes the proof of the lemma.
\end{proof}

\begin{figure}[ht]
\centering
\includegraphics[scale=1]{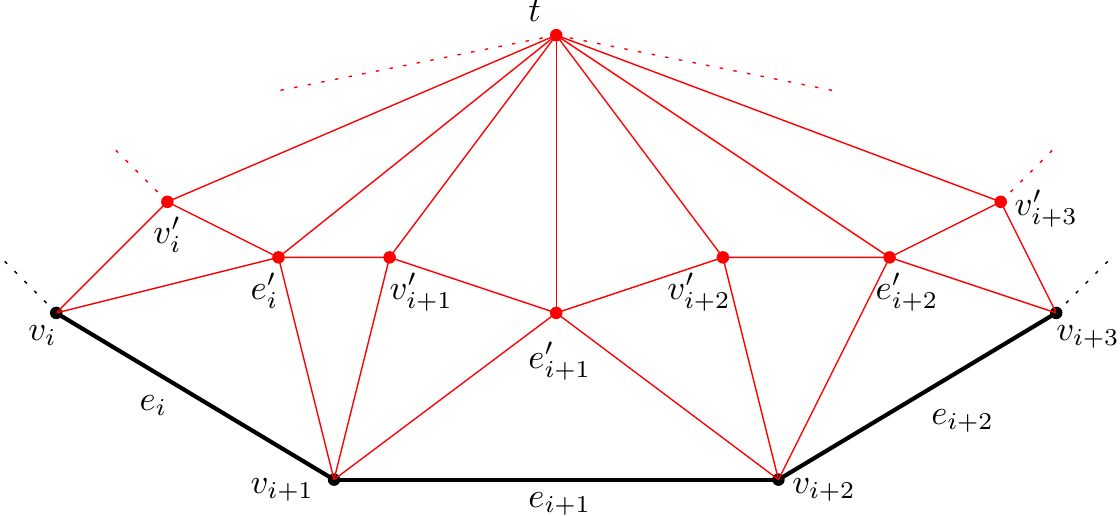}
\caption{Zoom on the new connections.}
\label{fig:tf24c_gadget}
\end{figure}

\bibliographystyle{plain}
\bibliography{biblio.bib}

\begin{thebibliography}{10}

\bibitem{AFjgaa15}
N.~Aerts and S.~Felsner.
\newblock {Vertex Contact Representations of Paths on a Grid}.
\newblock {\em Journal of Graph Algorithms and Applications}, 19(3):817 -- 849,
  2015.

\bibitem{asinowski2012vertex}
A.~Asinowski, E.~Cohen, M.C. Golumbic, V.~Limouzy, M.~Lipshteyn, and M.~Stern.
\newblock Vertex intersection graphs of paths on a grid.
\newblock {\em J. Graph Algorithms Appl.}, 16(2):129--150, 2012.

\bibitem{Ben91}
I.~Ben-Arroyo~Hartman, I.~Newman, and R.~Ziv.
\newblock {On grid intersection graphs}.
\newblock {\em Discret. Math.}, 87:41--52, 1991.

\bibitem{BDarxiv15}
T.~Biedl and M.~Derka.
\newblock {1-String B1-VPG Representations of Planar Partial 3-Trees and Some
  Subclasses}.
\newblock {\em ArXiv e-prints}, 2015.

\bibitem{BDjocg16}
T.~Biedl and M.~Derka.
\newblock {1-string $B_2$-VPG representation of planar graphs}.
\newblock {\em Journal of Computational Geometry}, 7(2), 2016.

\bibitem{BDjgaa16}
T.~Biedl and M.~Derka.
\newblock {The (3,1)-ordering for 4-connected planar triangulations}.
\newblock {\em Journal of Graph Algorithms and Applications}, 20(2):347--362,
  2016.

\bibitem{BiedlSOFSEM2017}
T.~Biedl and M.~Derka.
\newblock Order-preserving 1-string representations of planar graphs.
\newblock In {\em Proceedings of SOFSEM 2017}, pages 283--294, 2017.

\bibitem{BP17}
T.~Biedl and C.~Pennarun.
\newblock {4-connected graphs are in $B_3$-EPG}.
\newblock Work in preparation.

\bibitem{Cdam17max}
D.~Catanzaro, S.~Chaplick, S.~Felsner, B.V. Halld{\'o}rsson, M.M.
  Halld{\'o}rsson, T.~Hixon, and J.~Stacho.
\newblock Max point-tolerance graphs.
\newblock {\em Discrete Applied Mathematics}, 216:84--97, 2017.

\bibitem{chalopin2009every}
J.~Chalopin and D.~Gon{\c{c}}alves.
\newblock Every planar graph is the intersection graph of segments in the
  plane.
\newblock In {\em Proceedings of the forty-first annual ACM symposium on Theory
  of computing}, pages 631--638, 2009.

\bibitem{chalopin2010planar}
J.~Chalopin, D.~Gon{\c{c}}alves, and P.~Ochem.
\newblock Planar graphs have 1-string representations.
\newblock {\em Discrete \& Computational Geometry}, 43(3):626--647, 2010.

\bibitem{chaplick2012bend}
S.~Chaplick, V.~Jel{\'\i}nek, J.~Kratochv{\'\i}l, and T.~Vysko{\v{c}}il.
\newblock Bend-bounded path intersection graphs: Sausages, noodles, and waffles
  on a grill.
\newblock In {\em Graph-Theoretic Concepts in Computer Science}, pages
  274--285. Springer, 2012.

\bibitem{chaplick2013equilateral}
S.~Chaplick, S.G. Kobourov, and T.~Ueckerdt.
\newblock Equilateral {L}-contact graphs.
\newblock In {\em International Workshop on Graph-Theoretic Concepts in
  Computer Science}, pages 139--151. Springer Berlin Heidelberg, 2013.

\bibitem{chaplick2013planar}
S.~Chaplick and T.~Ueckerdt.
\newblock {P}lanar {G}raphs as {VPG}-{G}raphs.
\newblock {\em J. Graph Algorithms Appl.}, 17(4):475--494, 2013.

\bibitem{cohen2016posets}
E.~Cohen, M.C. Golumbic, W.T. Trotter, and R.~Wang.
\newblock Posets and {VPG} {G}raphs.
\newblock {\em Order}, 33(1):39--49, 2016.

\bibitem{Castro02}
N.~de~Castro, F.~Cobos, J.C. Dana, A.~Márquez, and M.~Noy.
\newblock {Triangle-free planar graphs as segment intersection graphs}.
\newblock {\em J. Graph Algorithms Appl.}, 6(1):7--26, 2002.

\bibitem{Fraysseix07}
H.~de~Fraysseix and P.~Ossona~de Mendez.
\newblock {Representations by contact and intersection of segments}.
\newblock {\em Algorithmica}, 47(4):453--463, 2007.

\bibitem{Fraysseix94}
H.~de~Fraysseix, P.~Ossona~de Mendez, and J.~Pach.
\newblock {Representation of planar graphs by segments}.
\newblock {\em Intuit. Geom. (Szeged, 1991), Colloq. Math. Soc. János Bolyai},
  63:109--117, 1994.

\bibitem{FOR94}
H.~de~Fraysseix, P.~Ossona~de Mendez, and P.~Rosenstiehl.
\newblock {On Triangle Contact Graphs}.
\newblock {\em Combinatorics, Probability and Computing}, 3:233--246, 1994.

\bibitem{ehrlich1976intersection}
G.~Ehrlich, S.~Even, and R.E. Tarjan.
\newblock Intersection graphs of curves in the plane.
\newblock {\em Journal of Combinatorial Theory, Series B}, 21(1):8--20, 1976.

\bibitem{Fdam16}
S.~Felsner, K.~Knauer, G.B. Mertzios, and T.~Ueckerdt.
\newblock Intersection graphs of {L}-shapes and segments in the plane.
\newblock {\em Discrete Applied Mathematics}, 206:48--55, 2016.

\bibitem{francis2016vpg}
M.C. Francis and A.~Lahiri.
\newblock {VPG} and {EPG} bend-numbers of {H}alin graphs.
\newblock {\em Discrete Applied Mathematics}, 215:95--105, 2016.

\bibitem{GHKK12}
E.R. Gansner, Y.~Hu, M.~Kaufmann, and S.G. Kobourov.
\newblock {Optimal Polygonal Representation of Planar Graphs}.
\newblock {\em Algorithmica}, 63(3):672--691, 2012.

\bibitem{GLP12}
D.~Gonçalves, B.~L\'ev\^eque, and A.~Pinlou.
\newblock {Triangle contact representations and duality}.
\newblock {\em Discrete and Computational Geometry}, 48:239--254, 2012.

\bibitem{Kapelle15}
B.~Kapelle.
\newblock {Kontact- und Schnittdarstellungen planarer Graphen}, 2015.

\bibitem{KKLS06}
M.~Kaufmann, J.~Kratochv\'il, K.A. Lehmann, and A.R. Subramanian.
\newblock {Max-tolerance graphs as intersection graphs: Cliques, cycles and
  recognition}.
\newblock {\em In Proc. SODA '06}, pages 832--841, 2006.

\bibitem{Kenyon04}
R.W. Kenyon and S.~Sheffield.
\newblock {Dimers, tilings and trees}.
\newblock {\em J. Comb. Theor. Ser. B}, 92:295--317, 2004.

\bibitem{kobourov2013combi}
S.~Kobourov, T.~Ueckerdt, and K.~Verbeek.
\newblock {Combinatorial and geometric properties of planar Laman graphs}.
\newblock In {\em Proceedings of the twenty-fourth annual ACM-SIAM symposium on
  Discrete algorithms (SODA 2013)}, pages 1668--1678. Society for Industrial
  and Applied Mathematics, 2013.

\bibitem{Koebe36}
P.~Koebe.
\newblock {Kontaktprobleme der konformen Abbildung}.
\newblock {\em Ber. Sächs. Akad. Wiss. Leipzig, Math. Phys. Kl.}, 88:141--164,
  1936.

\bibitem{Marxiv17}
S.~Mehrabi.
\newblock {Approximation Algorithms for Independence and Domination on
  $B_1$-VPG and $B_1$-EPG Graphs}.
\newblock {\em ArXiv e-prints}, 2017.

\bibitem{middendorf1992max}
M.~Middendorf and F.~Pfeiffer.
\newblock The max clique problem in classes of string-graphs.
\newblock {\em Discrete mathematics}, 108(1-3):365--372, 1992.

\bibitem{2017arXiv170301544R}
A.~{Reyan Ahmed}, F.~{De Luca}, S.~{Devkota}, A.~{Efrat}, M.~I. {Hossain},
  S.~{Kobourov}, J.~{Li}, S.~{Abida Salma}, and E.~{Welch}.
\newblock {L-Graphs and Monotone L-Graphs}.
\newblock {\em ArXiv e-prints}, 2017.

\bibitem{schein84}
E.R. Scheinerman.
\newblock {Intersection Classes and Multiple Intersection Parameters of
  Graphs}, PhD thesis, Princeton University, 1984.

\bibitem{Schramm93}
O.~Schramm.
\newblock {Square tilings with prescribed combinatorics}.
\newblock {\em Isr. J. Math.}, 84:97--118, 1993.

\bibitem{S90}
O.~Schramm.
\newblock {Combinatorically Prescribed Packings and Applications to Conformal
  and Quasiconformal Maps}.
\newblock {\em ArXiv e-prints}, 0709.0710, 2007.

\bibitem{Swg17}
H.~Schrezenmaier.
\newblock {Homothetic triangle contact representations}.
\newblock {\em Proceedings of WG '17}, 2017.

\bibitem{T84}
C.~Thomassen.
\newblock {Plane representations of graphs}.
\newblock {\em Progress in graph theory (Bondy and Murty, eds.)}, pages
  336--342, 1984.

\bibitem{Whitney31}
H.~Whitney.
\newblock {A theorem on graphs}.
\newblock {\em Ann. Math.}, 32(2):378--390, 1931.

\end{thebibliography}

\end{document}